\magnification=\magstep1
\baselineskip=20pt
\centerline{After HETE}
\bigskip
\centerline{J. I. Katz}
\medskip
\centerline{Racah Institute of Physics, Hebrew University, Jerusalem, Israel}
\centerline{and}
\centerline{Department of Physics and McDonnell Center for the Space Sciences}
\centerline{Washington University, St. Louis, Mo.\footnote*{Permanent Address}}
\centerline{katz@wuphys.wustl.edu}
\bigskip
\centerline{Abstract}
\medskip
The outstanding task of gamma-ray burst astronomy is to test the hypothesis
that they are at cosmological distances.  This can be done by determining 
the coordinates of at least a few bursts to $\sim 15^{\prime\prime}$ or 
better.  If they are in distant galaxies, the next task is to determine from
which component of the galaxies the bursts originate; sub-arcsecond positions 
would answer this question.  I outline ground-based systems which can 
accomplish these tasks, given BATSE/BACODINE coordinates and conservative
assumptions about visible counterparts; these systems would be a 
ground-based upgrade to GRO which would accomplish the chief objectives of 
HETE.
\vfil
\eject
The loss of HETE on November 4, 1996, following its failure to separate from
its booster after launch, forces reconsideration of the future of gamma-ray
burst (GRB) astronomy.  We should ask again what are the important questions,
and which observations and experiments may answer them.  In addition,
because space flights are few and far between (HETE's unsuccessful launch 
occurred nearly 15 years after its conception), we should try to anticipate 
the following round of questions, and to design instruments which can answer
them too.

There are two reasons for attempting to detect GRB in visible or 
ultraviolet light: to measure their energy spectrum, and to obtain accurate
coordinates.  The energy spectrum constrains emission mechanisms and
theoretical models.  Visible and ultraviolet telescopes have the ability
to measure source coordinates accurately, often to better than an arc second.
Such accurate coordinates are necessary for conclusive identifications with 
other astronomical objects.  If accurate coordinates lead only to blank 
fields, that itself would be an important result.  Nonetheless, there is 
broad agreement that obtaining accurate coordinates is the most promising 
attack on the GRB problem, and that they can best be obtained from visible 
or UV counterparts.

This line of reasoning led to the conception of HETE in the early 1980's.
It was then believed$^{1,2}$ that
GRB were accompanied by visible light transients bright enough to be seen
by the naked eye (supposing a naked eye looking in the right direction at
the right time; the difficulty is that the implied duty cycle of bright GRB 
is $\sim 10^{-7}$/sterad).  However, these early suggestions of bright
visible counterparts to GRB are now widely viewed with skepticism.  Doubts
have been raised$^3$ about the original images and the subject remains
controversial$^4$.  Ground-based instruments like the Explosive Transient
Camera$^5$ and GROSCE$^6$ have found no visible transients coincident and
simultaneous with GRB, and may also be inconsistent with the suggested
archival transients.  In addition, an association of the suggested archival
transients with GRB would require a repetition time of about a year, while 
the BATSE data exclude repetition times less than $\approx 100$ years$^7$
(if short repetition times were found only
for visible light and not for gamma-ray emission, then an instrument
like HETE would lack its required gamma-ray trigger).  Such bright visible
events were expected when it was believed GRB originated from neutron stars
at distances $\sim 100$ pc, some of which might have nondegenerate binary 
companions and therefore show a strong reflection effect, but this 
distance scale has now been excluded.

HETE's UV detector$^8$ had a limiting magnitude ($\lambda = 2500$\AA) of
about 7.  In the absence of strong empirical evidence for visible
counterparts to GRB of naked eye brightness, such an instrument is not well
suited to the problem: there was little reason to expect the UV detector
on HETE to observe anything at all.  In this respect HETE was obsolete
before it was launched.  Without the likelihood of UV detections, HETE
would neither determine the extension of the GRB spectrum over four decades
of photon energy nor find accurate coordinates.  It would have been left 
with the ability to determine coordinates to $\sim 20^\prime$ accuracy from 
its X-ray detector, roughly ten times better than BATSE.  These coordinates
could be very useful for directing simultaneous ground-based observations, 
but I will argue that the same objectives can be achieved with BATSE 
coordinates.  HETE's $\sim 20^\prime$ coordinates would probably be
of little value for follow-up observations, because more accurate IPN
coordinates have yielded no significant identifications of classical GRB.

With no generally accepted direct evidence for the visible (or UV) brightness
of GRB, either during outburst or afterwards, how do we decide how sensitive
an instrument must be?  Theoretical arguments$^{9,10}$ (necessarily 
model-dependent) suggest that below their soft gamma-ray peak the 
instantaneous spectral flux density of GRB approaches the asymptotic
synchrotron spectrum
$$F_\nu \propto \nu^{1/3}. \eqno(1)$$  
This is approximately consistent with observations in the range 20--100 KeV,
and may be a fair empirical extrapolation even if the theoretical models are
rejected.  Self absorption is not expected to be important above microwave
frequencies$^{11}$.  Then the ratio of broad-band fluxes 
$$F_{opt}/F_\gamma \sim 2 \times 10^{-7}. \eqno(2)$$  
For the ``burst of the month''$^{12}$ $F_\gamma \sim 10^{-5}$ erg/cm$^2$sec,
corresponding to a visible counterpart of 18th magnitude and explaining why
searches for them have so far been unsuccessful.  Some arguments$^9$ suggest
that the simultaneous visible emission will be followed by delayed emission 
several magnitudes brighter, and the instantaneous asymptote may not 
be reached in measurements of finite duration; in some cases $F_\nu$ appears
to vary as a significantly lower power of $\nu$ than 1/3 in the 20--100 KeV
range, although there is no evidence that a lower exponent applies at
photon energies below 20 KeV.  Each of these possibilities implies a visible
brightness greater than that given by Equation (2).  Still, a conservative 
instrument designer will consider 18th magnitude a prudent goal, and 
instruments not capable of reaching this limit may risk seeing nothing.

What should our scientific objectives be?  Most astrophysicists now believe
that GRB originate at cosmological distances, with the brighter ``bursts of
the month'' probably at redshifts $\sim 0.2$--0.5.  Testing this hypothesis
is therefore the most important task of GRB astronomy.  It is likely (but
unproven) that if GRB are at cosmological distances they are associated
with galaxies.  At these redshifts, galaxies$^{13}$ with visible magnitudes
21--22 have a density on the sky of roughly 1/arcmin$^2$.  Hence GRB
coordinates of accuracy $\ll 1^\prime$ are sufficient to test the hypothesis
that they are associated with distant galaxies; if the hypothesis is correct
there will be dramatic pictures of small GRB error boxes neatly enclosing
galaxies.  A few such images would convince even die-hard skeptics.  Less 
accurate coordinates would, at best, lead to controversial statistical 
arguments.  If GRB are not associated with distant galaxies, then they may 
be associated with some unexpected faint objects.  The more accurate the 
coordinates, the stronger the case which could be made for such an 
association.

Suppose GRB do turn out to come from distant galaxies.  The next question 
to ask is where in their host galaxies they are found: in the discs (if the
galaxies are spiral), extended halos, central bulges or elliptical
components, or in the nuclei?  The answer would tell us a great deal about
the origin and dynamical evolution of the objects which produce GRB.
Coordinates of $0.^{\prime\prime}2$ accuracy would answer this question (at
$z = 0.3$ an angle $0.^{\prime\prime}2$ corresponds to 1.5 Kpc). and this 
is a natural goal of GRB astronomy.

HETE could not have answered these questions by itself.  Its UV coordinates 
of $\sim 20^{\prime\prime}$ accuracy would probably have been accurate 
enough to decide if GRB come from distant galaxies, but it was unlikely to 
have detected GRB in the UV.  Its X-ray coordinates of $\sim 20^\prime$ 
accuracy could have enabled ground-based observatories to detect visible
counterparts of GRB and from them to determine accurate coordinates.
However, it is possible to achieve this objective using only BATSE's rougher
coordinates distributed in real time by its BACODINE network.  The technology
required is used in instruments such as ETC and GROSCE, and can work even
if only BATSE coordinates are available.

It is not necessary to locate every GRB.  It would be sufficient to produce
one accurate set of coordinates per year, so I assume a ``burst of
the month'' in order to allow for various inefficiencies and duty factors
less than unity, such as daylight, moonlight, bad weather at the ground
stations, Earth occultation, equipment limitations and failures, and bursts 
in regions of heavy extinction or 
more than $1\sigma$ from their nominal coordinates.  The coordinates are 
transmitted by BACODINE in less than 7 seconds$^{14}$.  Roughly half of all 
bursts have durations of 20 seconds or more, so there is ample time for the 
ground stations to respond (short GRB, which may be a class of events 
distinct from long GRB, cannot be studied with any system which depends on 
measurement and transmission of coordinates and slewing of a telescope).  

The ground stations each consist of a system of two rapidly slewing 
telescopes.  The first telescope has two functions: to obtain coordinates of
$\sim 15^{\prime\prime}$ accuracy, and to measure the intensity history of
optical counterparts to GRB.  These coordinates are probably accurate enough
to decide if GRB are associated with distant galaxies.  They also permit
the second telescope to locate the GRB.  The second telescope serves only 
to determine sub-arc second coordinates.

The first telescope takes the place of the UV telescope on HETE.  Suppose 
its active focal plane consists of four $1024 \times 1024$ CCDs with $25\mu$
pixels.  Each CCD views a $4^\circ$ square patch of sky, the four together
covering a square $8^\circ$ on a side, which includes the $1 \sigma$ error 
circle$^{14}$ for BACODINE coordinates of the bright ``bursts of the month'',
so that more than two thirds of these bursts will be within the
field of view.  Such a telescope has an aperture of 15 inches at $f/1$, and 
its pixels have angular size $15^{\prime\prime}$.  With plausible estimates 
for the band pass, source spectrum and quantum efficiency the count rate for
an 18th magnitude object is $\sim 100$/sec.  A dark (moonless) sky has a 
brightness of about 16th magnitude per $15^{\prime\prime}$ square, or $\sim 
600$ counts/sec, permitting $4\sigma$ detection in one second of 
integration.  Akerlof, {\it et al.}$^{15}$ give similar estimates.  
Discrimination against airglow may be achieved by spectral filtering, and 
against Zodiacal light by polarization filtering, but are probably not 
necessary.

By comparing successive one-second integrations it is possible to identify
all rapidly varying sources of visible light in the field of view and to
construct histories of intensity {\it vs.} time for each of them.  There
will be a background of transients resulting from meteors, aircraft and
satellite glints, which can readily be eliminated as is done by ETC.
Variable stars, especially flare stars, may also be detected.  Stellar 
flares have slower rise times and much slower decay times than most GRB (and
a more pronounced time asymmetry), and simple single-peaked profiles.  If 
the visible counterparts to GRB follow their gamma-ray time dependence then 
they may immediately be distinguished from stellar flares.  Flare stars also
differ from GRB in having quiescent counterparts and repetition rates $\sim 
1$/hr.  The intensity histories of GRB serve to discriminate them from other
signals but are also of intrinsic interest.

The second telescope slews to the coordinates provided by the first 
telescope.  If more than one transient is detected it must survey each in 
turn, emphasizing the desirability of discriminating against transients 
other than GRB.  With a much smaller field of view it can have a much finer 
angular resolution; for example, a $1024 \times 1024$ element CCD in its focal
plane would provide $0.^{\prime\prime}2$ pixels over a $200^{\prime\prime}$
field of view.  The effect of sky brightness is negligible for 18th 
magnitude GRB counterparts at this angular resolution.  This second telescope
could have an aperture as small as that of the first telescope, but a larger
aperture is desirable to increase the signal to noise ratio and to provide 
images of faint galaxies and stars (for differential astrometry) within its
limited field of view.

By finding the centroid of a $1^{\prime\prime}$ seeing disc it will be 
possible to locate the GRB counterpart (with respect to nearby stars) to a
fraction of an arc second, as is conventionally done in astrometry.  If 
there were no readout noise or other sources of error beyond ideal counting 
statistics then an 18th magnitude object could be located to about 
$0.^{\prime\prime}2$ in a second of integration, and, more realistically,
to this or better accuracy by integrating throughout its duration.  These 
astrometric coordinates can be compared to follow-up observations by larger
telescopes of the same field to determine if there are galaxies at the 
coordinates of the GRB.  If so, HST observations could determine where in 
their host galaxies GRB are found.

As with all ground-based systems, constraints of moonlight, weather, 
daylight, zenith angle, Galactic extinction and equipment reliability imply 
an effective availability $\sim 3\%$.  It will be necessary to construct 
such telescope systems at a number of sites around the world, distributed in
longitude and latitude, in order to achieve one detection per year of a 
``burst of the month''.

This system uses much of the same technology as ETC and GROSCE.  Its 
advantage over staring systems like ETC is that gamma-ray cuing acts as a 
very strong temporal filter (discrimination $\sim 10^5$) against spurious 
events, and even the rough coordinates provided by BATSE/BACODINE provide 
a $\sim$ hundred-fold reduction in the solid angle viewed.  This further
reduces the spurious event rate.  More important, it increases sensitivity 
(in inverse proportion to the solid angle which must be accepted) by 
permitting the use of a much larger aperture.  GROSCE II$^6$ shares these 
advantages, and resembles the first telescope described here, although 
I believe it prudent to plan a system which can reach 18th magnitude.  
The combined system can accomplish all the objectives of HETE
and its ground-based adjuncts, using the comparatively crude BATSE/BACODINE
coordinates in place of the more accurate HETE X-ray coordinates and without 
making the optimistic assumptions about UV brightness required by HETE.

I thank S. Barthelmy, S. Kobayashi and T. Piran for discussions, Washington
University for sabbatical leave, the Hebrew University for hospitality and 
a Forchheimer Fellowship and NASA NAG 52682 and NSF AST 94-16904 for support.
\vfil
\eject
\centerline{References}
\medskip
\item{1.} Schaefer, B. E., {\it Nature} {\bf 294}, 722 (1981).
\item{2.} Schaefer, B. E., {\it et al.} {\it Ap. J. Lett.} {\bf 286}, L1 
(1984).
\item{3.} \.Zytkow, A. {\it Ap. J.} {\bf 359}, 138 (1990).
\item{4.} Hudec, R. {\it Proc. 29th ESLAB Symp.}, {\it Ap. Sp. Sci.} in press
(1996).
\item{5.} Vanderspek, R., Krimm, H. A. and Ricker, G. R. {\it Proc. 29th
ESLAB Symp.}, {\it Ap. Sp. Sci.} in press (1996).
\item{6.} Akerlof, C. {\it et al.} {\it Proc. 29th ESLAB Symp.}, {\it Ap. Sp.
Sci.} in press (1996).
\item{7.} Tegmark, M., Hartmann, D. H., Briggs, M. S., Hakkila, J. and
Meegan, C. A. {\it Ap. J.} {\bf 466}, 757 (1996).
\item{8.} Vanderspek, R., {\it et al.} {\it Proc. 29th ESLAB Symp.}, {\it Ap.
Sp. Sci.} in press (1996).
\item{9.} Katz, J. I. {\it Ap. J. Lett.} {\bf 432}, L107 (1994).
\item{10.} Tavani, M. {\it Phys. Rev. Lett.} {\bf 76}, 3478 (1996).
\item{11.} Katz, J. I. {\it Ap. J.} {\bf 422}, 248 (1994).
\item{12.} Fishman, G. J. and Meegan, C. A. {\it Ann. Rev. Astron. Ap.} {\bf
33}, 415 (1995), Fig. 13.
\item{13.} Cohen, J. G., {\it et al.} {\it Ap. J.} {\bf 471}, L5 (1996).
\item{14.} Barthelmy, S. D., {\it et al.} {\it Proc. 29th ESLAB Symp.}, {\it
Ap. Sp. Sci.} in press (1996).
\item{15.} Akerlof, C. {\it et al.} {\it Gamma Ray Bursts, Second Workshop,
Huntsville} eds. G. J. Fishman, J. J. Brainerd, K. Hurley (AIP Conf. Proc.
{\bf 307}), p. 633 (1994).
\vfil
\eject
\bye
\end